\documentclass[fleqn,10pt]{wlscirep}
\usepackage[utf8]{inputenc}
\usepackage[T1]{fontenc}
\usepackage[switch]{lineno}
\usepackage{textgreek}
\usepackage{gensymb}
\usepackage{subcaption}
\usepackage{ulem}
\usepackage{rotating}
\usepackage{upgreek}

\title{Attosecond gamma-ray flashes and electron-positron pairs in dyadic laser interaction with micro-wire}

\author[1,*]{Prokopis Hadjisolomou}
\author[1]{Tae Moon Jeong}
\author[1]{Petr Valenta}
\author[1]{Alexander J. Macleod}
\author[1]{Rashid Shaisultanov}
\author[2]{Christopher P. Ridgers}
\author[1,3]{Sergei V. Bulanov}
\affil[1]{ELI Beamlines Facility, Extreme Light Infrastructure ERIC, Za Radnicí 835, 25241 Dolní Břežany, Czech Republic}
\affil[2]{York Plasma Institute, Department of Physics, University of York, Heslington, York, North Yorkshire YO10 5DD, UK}
\affil[3]{National Institutes for Quantum Science and Technology (QST), Kansai Photon Science Institute, 8-1-7 Umemidai, Kizugawa, Kyoto 619-0215, Japan}
\affil[*]{prokopis.hadjisolomou@eli-beams.eu}

\begin{abstract}
The interaction of an ultra-intense laser with matter is an efficient source of high-energy particles, with efforts directed towards narrowing the divergence and simultaneously increasing the brightness. In this paper we report on emission of highly collimated, ultrabright, attosecond \textgamma-photons and generation of dense electron-positron pairs via a tunable particle generation scheme which utilizes the interaction of two high-power lasers with a thin wire target. Irradiating the target with a radially polarized laser pulse first produces a series of high charge, short duration, electron bunches with low transverse momentum. These electron bunches subsequently collide with a counter-propagating high intensity laser. Depending on the intensity of the counter-propagating laser, the scheme generates highly collimated ultra-bright GeV-level \textgamma-beams and/or electron-positron plasma of solid density level.
\end{abstract}

\begin{document}

\flushbottom
\maketitle

\thispagestyle{empty}

\twocolumn

Extragalactic \textgamma-ray bursts are the most energetic and highest luminosity events observed in astrophysics since the Big Bang \cite{2005_PiranT}. Within a few seconds, the energy of the emitted \textgamma-photon radiation reaches the $10^{54} \kern0.2em \mathrm{erg}$ level. This raises one of the most intriguing questions in astrophysics: how to explain the mechanism by which \textgamma-ray bursts are generated in space. According to current concepts \cite{1992_MeszarosP, 2001_RuffiniR}, powerful \textgamma-ray bursts occur when stars collide in binary star systems, followed by the collapse of one of the stars and the formation of black holes. As a result of this process, relativistic jets of lepton-hadron plasma are formed, described within the framework of the fireball paradigm \cite{1992_MeszarosP}, the generation of electron-positron ($e^{-}e^{+}$) pairs, the development of plasma instabilities, magnetic reconnection and collisionless shock waves accelerating charged particles to high energies. Gamma-ray bursts are characterized by an extremely high efficiency of converting the energy of charged particles into \textgamma-ray energy \cite{2006_FanY}. Although many hypotheses have been proposed to explain the origin of \textgamma-ray bursts, there is no doubt that an extremely strong electromagnetic field accelerates high-energy electrons from the plasma, which then interact with the strong electromagnetic field, emitting high-energy photons. One of the versions of that scenario is known as the synchrotron self-Compton radiation of high-energy photons \cite{1979_GouldRJ}.

Extremely high efficiency high-power \textgamma-ray flash can be generated in the multi-Petawatt power laser-plasma interaction \cite{2012_NakamuraT, 2012_RidgersCP, 2018_LezhninKV, 2023_HadjisolomouP}. A \textgamma-ray burst in a laser plasma is also the result of a sequence of processes in which high-energy electrons in the plasma are accelerated by a strong electromagnetic field and then interact with the strong electromagnetic field, emitting high-energy photons. It can be considered as an analogue of the synchrotron self-Compton radiation in an ultra-high-intensity electromagnetic field. The \textgamma flash in laser plasma is accompanied by the creation of an $e^{-}e^{+}$ pair plasma. 

%Let us imagine a scenario where an electromagnetic field inputs to a `black box' and outputs an enormous population of \textgamma-photons and $e^{-}e^{+}$ pairs; as seen below, our paper realizes how such a scenario can be materialized. The interaction of matter with electromagnetic field occurs in astrophysical scales, where \textgamma-photon emission has been observed in the form of \textgamma-ray bursts \cite{2001_RuffiniR, 2001_RuffiniRb, 2019_AmiriM, 2020_MarcoteB}. Moreover, the unique plasma features achieved by our setup in comparison to other schemes falls into the strong-field quantum electrodynamics environment where $e^{-}e^{+}$ pairs exhibit collective effects. Quantum electrodynamics plasmas lay on a strong theoretical framework, of which the experimental verification faces challenges on how to achieve such a dense $e^{-}e^{+}$ pair plasma \cite{2020_ZhangP, 2023_ChenH, 2024_ArrowsmithCD}. Such dense plasmas can occur in the viscinity of neutron stars \cite{2022_PhilippovA} and/or black holes \cite{2016_BambiC} and have recently been achieved in the laboratory with an $e^{-}e^{+}$ pair density of $\sim 10^{18} \kern0.2em \mathrm{m^{-3}}$ with the aim of testing the microphysics of astrophysical observations \cite{2024_ArrowsmithCD}. We hope our proposed scheme, predicting an $e^{-}e^{+}$ pair density more than ten orders of magnitude higher, to pave the way on generation of such plasma environments and illuminate novel insights into fundamental physics.

High-energy photon and electron/positron beams have found important uses across a range of disciplines in modern science. 
A non-exhaustive list of their applications include radiotherapy \cite{1997_weeksKJ, 2002_Papiez}, photonuclear fission \cite{2000_CowanTE, 2003_SchwoererH, 2004_NedorezovVG}, study of shock-waves \cite{2008_RavasioA, 2017_AntonelliL}, materials science \cite{2013_EliassonB, 2009_Migley, 2005_King}, nuclear physics \cite{2014_TarbertCM, 1987_Foris}, neutron sources \cite{2014_PomerantzI}, astrophysical studies \cite{1992_ReesMJ, 2015_BulanovSV, 2018_PhilippovAA, 2021_Takabe}, study of fundamental physics \cite{1998_Telnov, 2012_Tajima, 2023_FedotovA} and positron generation \cite{2000_GahnC, 2015_SarriG, 2022_KolenatyD, 2023_MacLeodAJ}. Typically, generating high-energy particle beams requires large scale facilities and technology such as radio frequency accelerators. This can limit their availability for certain applications, motivating a growing interest in alternative sources.

The invention of the chirped pulse amplification technique\cite{1985_StricklandD} enabled the development of high power laser systems of pettawatt class, which have since become attractive candidates for particle production technologies\cite{2003_Bingham,2008_Malka,2013_Hooker,2020_Tajima}.
Multi-petawatt laser facilities are now a reality with demonstration of the $10 \kern0.2em \mathrm{PW}$ level in ELI-NP \cite{2020_TanakaKA, 2022_RadierC}, while a $10 \kern0.2em \mathrm{PW}$ laser system with an order of magnitude higher energy is soon expected at ELI-Beamlines \cite{2019_DansonC}. Moreover, focusing the laser beam at a spot of $1.1 \kern0.2em \mathrm{\upmu m}$ allowed surpassing the intensity level of $10^{23} \kern0.2em \mathrm{W cm^{-2}}$ with a $4 \kern0.2em \mathrm{PW}$ laser \cite{2021_YoonJ}. The development of a $100 \kern0.2em \mathrm{PW}$ laser is ongoing \cite{2020_Peng, 2021_LiZ}, while the dual-beam $25 \kern0.2em \mathrm{PW}$ EP-OPAL laser system \cite{2022_DiPiazzaA, 2023_ZuegelJ} will allow more complex laser-target interaction setups.

The interaction of an ultra-intense laser ($\sim 10^{22} \kern0.2em \mathrm{W cm^{-2}}$) with matter results in the creation of energetic particle populations \cite{2006_MourouGA}, namely electrons, positrons and \textgamma-photons. In particular, irradiating a solid density target with a high-power laser has been shown to be a particularly simple and efficient scheme for particle acceleration\cite{2022_GonoskovA}. This results in a \textgamma-ray flash \cite{2012_NakamuraT, 2012_RidgersCP, 2023_HadjisolomouP}, where high-energy \textgamma-photons are emitted via multiphoton Compton scattering \cite{1964_NikishovAI, 1964_BrownLS, 1964_GoldmanII, 1985_RitusVI, 2012_RidgersCP, 2013_RidgersCP, 2022_GonoskovA} , and the creation of high charge electron and positron beams via the multiphoton Breit-Wheeler pair-production \cite{1962_ReissHR,1964_NikishovAI,1966_YakovlevVP}. While this scheme produces high particle yields and efficiently converts laser energy into total particle energy, some of the beam properties limit their usefulness for certain applications. In particular, when the high-power laser is linearly polarised both the \textgamma-photons and electron-positron beams are emitted with large divergences, leading to low brightness.

In this paper we introduce a novel laser-matter interaction scheme for production of high-brightness strongly collimated \textgamma-ray beams and/or high-charge and high-density bunches of $e^{-}e^{+}$ pairs which surpass the solid density level. The beams in each case appear in the form of a series of attosecond duration bunches with low-divergence. The proposed scheme is based on the interaction of two ultraintense lasers with a wire target. The overall process can be viewed as three step model, namely injection (first) , boosting (second) and collision (third) stage. Initially, a radially polarized (RP) laser \cite{2018_JeongTM} interacts with the wire target, ejecting localized electron bunches and accelerating them \cite{2007_KarmakarA, 2013_VarinC, 2020_ZaimN, 2024_PowellJ, 2024_SunT}. Once the electrons gain sufficient energy from the RP laser, a linearly polarised (LP) laser collides head-on with them yielding predominantly either a \textgamma-ray flash or a large number of $e^{-}e^{+}$ pairs, depending on the focusing conditions of the LP laser.
%

%Our scheme can be thought as an electromagnetic field input to a `black box' resulting in the output an enormous population of \textgamma-photons and $e^{-}e^{+}$ pairs, in close resemblance of processes occurring in astrophysical environments, and is linked in particular with generation of fast radio bursts \cite{2019_AmiriM, 2020_MarcoteB}. The unique plasma features achieved by out setup in comparison to other schemes falls into the strong-field quantum electrodynamics environment where $e^{-}e^{+}$ pairs exhibit collective effects. Quantum electrodynamics plasmas lay on a strong theoretical framework, of which the experimental verification faces challenges on how to achieve such a dense $e^{-}e^{+}$ pair plasma \cite{2020_ZhangP, 2023_ChenH}. We hope our proposed scheme to pave the way on generation of such plasma environments and illuminate novel insights into fundamental physics.

\begin{figure*}[ht!]
\centering
\includegraphics[width=0.7\linewidth]{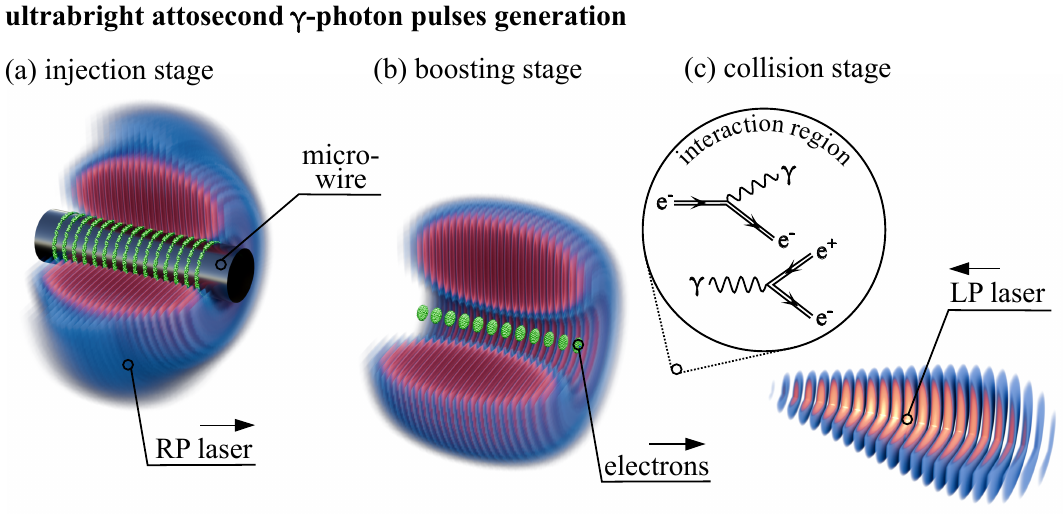}
\caption{Illustrative interaction setup. {\bf{(a)}} An $\mathrm{f/4}$-RP laser interacts with a lithium wire target, resulting in axial electron emission. {\bf{(b)}} The emitted electrons are accelerated up to the GeV energy level. {\bf{(c)}} The emitted electrons collide with a counter-propagating LP laser. The laser-electron interaction results in a highly directional \textgamma-ray flash accompanied by dense $e^{-}e^{+}$ pair generation.}
\label{fig:fig1}
\end{figure*}

%%%%%%%%%%%%%%%%%%%%%%%%%%%%%%%%%%%%%%%%%%%%%%%%%%%%%%%%%%%%%%%%%%%%%%%%%%%%
%%%%%%%%%%%%%%%%%%%%%%%%%%%%%%%%%%%%%%%%%%%%%%%%%%%%%%%%%%%%%%%%%%%%%%%%%%%%

\section*{Scheme Description}

The staged interaction process is depicted in Fig. \ref{fig:fig1}. At the injection stage (Fig. \ref{fig:fig1}(a) ), a RP laser focused by a parabola of f-number (the ratio of focal length to laser beam diameter) of 4 ($\mathrm{f/4}$-RP laser) interacts with a micro-wire target. Alternative proposed schemes using two or more lasers are based on symmetric laser-target interaction scenarios \cite{2015_LuoW, 2015_ZhangP, 2016_GrismayerT, 2016_VranicM, 2017_GongZ, 2017_CrebotI, 2017_LiHZb, 2020_YasenN}. Although interaction of strong lasers with micro/nano-thick wires has been previously studied \cite{2017_BargstenC, 2017_LiHZ,  2018_WangWM, 2021_ZhangL}, according to our knowledge none of these works employed RP mode, which is crucial for our scheme. The target is a cylinder of $\lambda$ radius and $4 \lambda$ length, where $\lambda$ is the laser wavelength. The target density corresponds to a lithium micro-wire, at an electron number density of $n_e = 1.39 \times 10^{29} \kern0.2em \mathrm{m^{-3}}$. Lithium, belonging in the alkali metals group, is the solid with lowest density, thus allowing easy penetration of the laser field in its volume; its skin depth has a relatively large value of $c/\omega_p \approx 14.3 \kern0.2em \mathrm{nm}$, where $\omega_p = \sqrt{n_e e^2 / (\varepsilon_0 m_e)}$ is the plasma frequency, $m_e$ is the electron mass, $e$ is the elementary charge and $\varepsilon_0$ is the vacuum permittivity. The chosen target density is near-optimal for the $25 \kern0.2em \mathrm{PW}$, $25 \kern0.2em \mathrm{fs}$ laser pulses used here. Lower density (foam) targets \cite{2016_StarkDJ} cannot be used in the micro-wire target geometry due to localised density spikes, forbidding fabrication in the form of a micro-wire. Higher density targets allow lower field penetration in the target volume, resulting in less electron ejection after a threshold. Notably, wire targets are geometrically opposite to hollow cylindrical (or conical \cite{2018_ZhuXLb, 2019_ZhuXL}) targets, extensively studied for attosecond electron bunches emission \cite{2004_NaumovaN, 2004_NaumovaNM} and \textgamma-photon emission \cite{2020_WangT, 2021_RinderknechtHG}. Attosecond electron modulations are also observed in `two-dimensional' wires \cite{2006_MaYY, 2024_ShenX}.

In this work we used the three-dimensional (3D) EPOCH \cite{2015_ArberTD} particle-in-cell (PIC) code, as described in the Methods section. The RP laser field is imported into the simulation box at the edge of its Rayleigh range. Unlike LP lasers which focus to a central peak, RP lasers focus to a ring; with only exception being the $\lambda^3$ regime \cite{2002_MourouG}. If the LP laser interacts with a wire of diameter larger than the focal spot diameter, then the electrons pile up near the target front surface forming a steep density gradient, in a similar way a flat-foil target interacts. On the other hand, if the wire dimensions are significantly smaller than the laser focal spot, then electrons are accelerated by the laser \cite{Wang1998, Zhy1998, Wang2001, Braenzel2017}, but in the expense of volumetric reduction of the electron number, thus making the scheme not efficient. RP lasers bypass both of the aforementioned limitations, allowing laser acceleration of electrons in high density bunches. However, the wire radius should be small enough not to modify the propagation of the accelerating field; for the $\mathrm{f/4}$-RP laser used, a target radius of $\lambda$ is near-optimal. The wire length is also important, as a long wire results in reduced energy of the ejected electrons due to attractive Coulomb forces. For our simulation parameters, a wire of $4 \lambda$ length is used, as it results in high ejected electron energy.

%In real experiments, the laser pulse is difficult to accurately incident on a micro-wire target, and may be misaligned with the electron beam. Thus, a more complex target configuration could implement either a conical \cite{2018_ZhuXLb} or cylindrical \cite{2019_ZhuXL} target to guide the laser pulse to the micro-wire target. Such guiding structures interacting with intense lasers result in bright \textgamma-rays and dense $e^{-}e^{+}$ pairs but with a divergence angle an order higher than our micro-wire target.

\begin{figure*}[ht]
\begin{subfigure}{1.0\textwidth}
  \centering
  \begin{sideways} injection, $\kern0.5em t = 50 \kern0.2em \mathrm{fs}$ \end{sideways} \fbox{\includegraphics[width=0.96\linewidth]{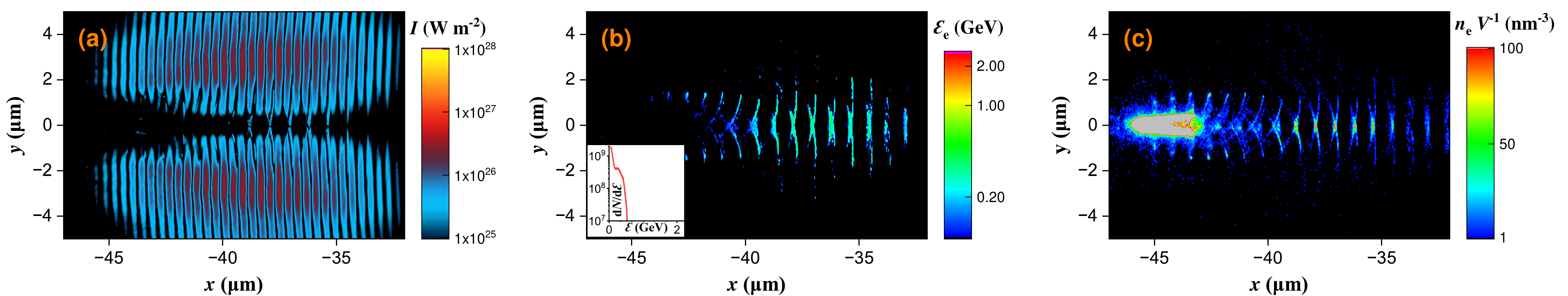}}
\end{subfigure}
\begin{subfigure}{1.0\textwidth}
  \centering
  \begin{sideways} boosting, $\kern0.5em t = 144 \kern0.2em \mathrm{fs}$ \end{sideways} \fbox{\includegraphics[width=0.96\linewidth]{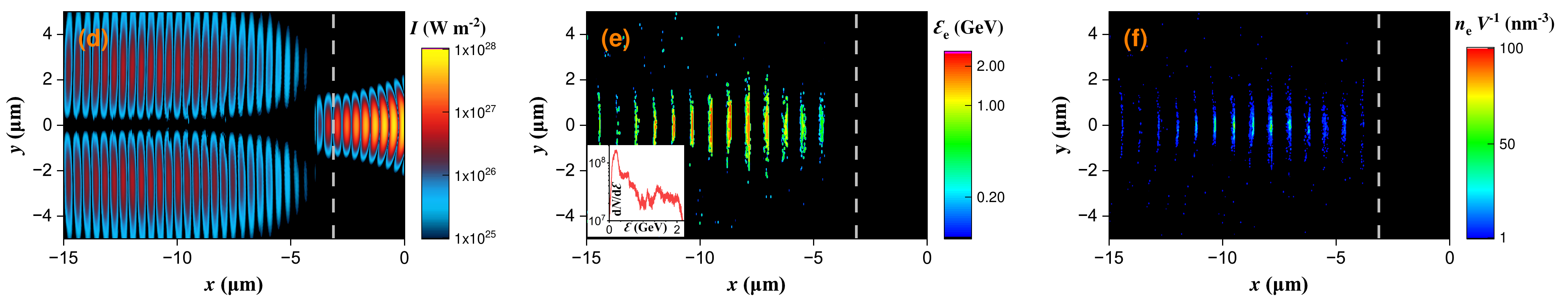}}
\end{subfigure}
\begin{subfigure}{1.0\textwidth}
  \centering
  \begin{sideways} collision, $\kern0.5em t = 170 \kern0.2em \mathrm{fs}$ \end{sideways} \fbox{\includegraphics[width=0.96\linewidth]{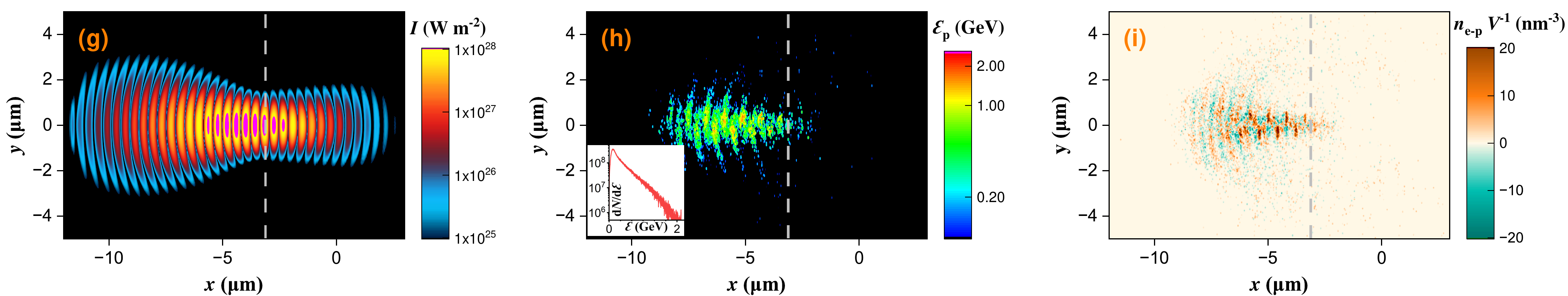}}
\end{subfigure}
\caption{ {\bf{(a)}} Intensity, {\bf{(b)}} electron mean kinetic energy and {\bf{(c)}} electron number density, after the interaction of a $\mathrm{f/4}$-RP laser with a lithium micro-wire. {\bf{(d)}} Intensity, {\bf{(e)}} electron mean kinetic energy and {\bf{(f)}} electron number density, prior the interaction of collimated electron bunches (ejected from the micro-wire) with a LP counter-propagating laser. {\bf{(g)}} Intensity, {\bf{(h)}} positron mean kinetic energy and {\bf{(i)}} electron minus positron number density, after the laser-electron interaction. The LP colliding pulse shown, corresponds to an $\mathrm{f/1}$. The $\mathrm{f/4}$-RP and $\mathrm{f/1}$-LP lasers are of same energy, and the laser intensity is controlled by varying the focal spot. For Figs. \ref{fig:fig2}(b,e,h) the corresponding energy spectra are given as insets. }
\label{fig:fig2}
\end{figure*}

The simulation results are depicted in Fig. \ref{fig:fig2}. Figs. \ref{fig:fig2}(a-c), \ref{fig:fig2}(d-f) and \ref{fig:fig2}(g-i) correspond to the injection, boosting and collision stages respectively, where the three stages are schematically illustrated in Fig. \ref{fig:fig1}(a-c). The first figure column shows the laser intensity. For the injection and boosting stages, column two and three correspond to the electron mean kinetic energy and electron number density respectively. For the collision stage, the second column shows the generated positron mean energy and the third column shows the difference of electron to positron number densities.

Fig. \ref{fig:fig2}(a-c) is at $50 \kern0.2em \mathrm{fs}$ simulation time, sufficient for the RP laser to interact with the micro-wire. Fig. \ref{fig:fig2}(a) reveals that the laser field interacts with the wire target keeping it's propagation properties almost undisturbed, as the wire diameter is considerably smaller than the pulse diameter. Small field distortions are observed in the centre of the laser pulse due to electron bunching and acceleration. These electron bunches are dense enough for their emitted radiation to be observable on the laser intensity map.

In Fig. \ref{fig:fig2}(b) the electron bunches correspond to a spectrum with a cut-off energy of $\sim 0.6 \kern0.2em \mathrm{GeV}$. Fig. \ref{fig:fig2}(c) shows detachment of the electron bunches from the bulk target, moving along the laser propagation direction. The lower energy electrons are left in the target forming an exponential-like decaying distribution. For the case of the most profound bunch, by considering electrons within $\pm 1 \kern0.2em \mathrm{\upmu m}$ in y-direction and z-direction, we obtain a spectrum approximated by $8.32 \times 10^6 \exp \left[ -\left( (\mathcal{E}-1950) / 263.51 \right)^2 \right]$, where $\mathcal{E}$ is the electron energy in MeV, corresponding to $\sim 4 \times 10^9$ electrons.

%The continuity equation is $\partial_t n + \partial_x (n u_e) = 0$, where $n$ is the electron density and $u_e$ is the electron velocity. Let us define $\xi = x-\beta c t$, where $\beta$ is the phase velocity divided by $c$. We can rewrite $\partial_t n = \partial_\xi n \partial_t \xi = - \beta c \partial_\xi n$ and $\partial_x (n u_e) = \partial_\xi (n u_e) \partial_x \xi = \partial_\xi (n u_e) $. The continuity equation becomes $- \beta c \partial_\xi n + \partial_\xi (n u_e) = 0$, which has solution $(-\beta c + u_e) n = const. = -\beta c n_0$ where $n_0$ is the initial electron density. This can be written as $n = \beta c n_0 / (\beta c - u_e)$, where $u_e= c p_e (m_e^2 c^2 + p_e^2)^{-1/2}$ and $p_e$ is the electron momentum.

%The Lorentz equation is $dp_e/dt=-e E(\xi) = d\xi/dt dp_e/d\xi = (u_e-\beta) dp_e/d\xi$. By replacing $u_e$, this gives $d (c \sqrt{m_e^2 c^2 + p_e^2}-\beta p_e) / d\xi = -e E(\xi)$ integration of which gives $c \sqrt{m_e^2 c^2 + p_e^2} - \beta p_e = - e \int^\xi E(\eta) d\eta+const.$. By setting $E(\eta)= E_0 \cos(\eta)$ with $E_0$ the field amplitude, we get expression for $u_e(\xi)$ and then equation for $n(\xi)$ gives modulations.

In a RP laser, the longitudinal (x-direction) electric field component, $E_\parallel(\xi)$, in the center of the beam is finite and the magnetic field is zero \cite{2000_McDonaldKT, 2018_JeongTM}, as the field vectors projections are of same sign (in contrary to a LP laser). The diffraction angle is $\theta = w_0/z_0$, where $w_0$ is the beam waist and $z_0$ is the Rayleigh range, giving $E_\parallel(\xi)= \theta E$, where $E$ is the laser electric field. The electron motion follows the equation
\begin{equation}
\frac{dp_e}{dt} = \frac{dp_e}{d\xi} \frac{d\xi}{dt} = k (u_e-\beta c) \frac{dp_e}{d\xi} =-e E_\parallel(\xi) ,
\end{equation}
where $\xi = k (x- \beta c t)$, $k$ is the laser wave-vector, $\beta=u_{ph}/c$ is the phase velocity ($u_{ph}$) divided by the speed of light in free space ($c$), $u_e$ is the electron velocity and $p_e$ is the electron momentum. The phase velocity is $u_{ph}=c \sqrt{k_\parallel^2+k_\perp^2}/k_\parallel$, with $k_\perp = k \sin(\theta)$ and $k_\parallel = k \cos(\theta)$ being the perpendicular and longitudinal wave-vector components. Thus, $\beta=1/\cos{\theta}$. Since $u_e= c p_e / \sqrt{m_e^2 c^2 + p_e^2}$, we have
\begin{equation}
\begin{aligned}
k c \left( \frac{p_e}{\sqrt{m_e^2 c^2 + p_e^2}} -\beta \right) \frac{dp_e}{d\xi} = \\ k c \frac{d}{d\xi} \left( \sqrt{m_e^2 c^2 + p_e^2}-\beta p_e \right) = -e E_\parallel(\xi),
\end{aligned}
\end{equation}
integration of which gives
\begin{equation}
c \left( \sqrt{m_e^2 c^2 + p_e^2} - \beta p_e \right) = - \frac{e}{k} \int^\xi E_\parallel(\eta) d\eta+C \equiv \Phi(\xi),
\label{eq:3}
\end{equation}
where $C$ is an integration constant. From equation \ref{eq:3} one can get the expression of $u_e(\xi)$. By using the Lorentz factor, $\gamma=\sqrt{1+p_e^2/(m_e c)^2}$, in equation \ref{eq:3}, the electron energy for large $p_e$ is
\begin{equation}
\gamma m_e c^2 \approx \frac{\Phi(\xi)}{1-\beta},
\end{equation}

\begin{figure}[ht]
\centering
\includegraphics[width=1.0\linewidth]{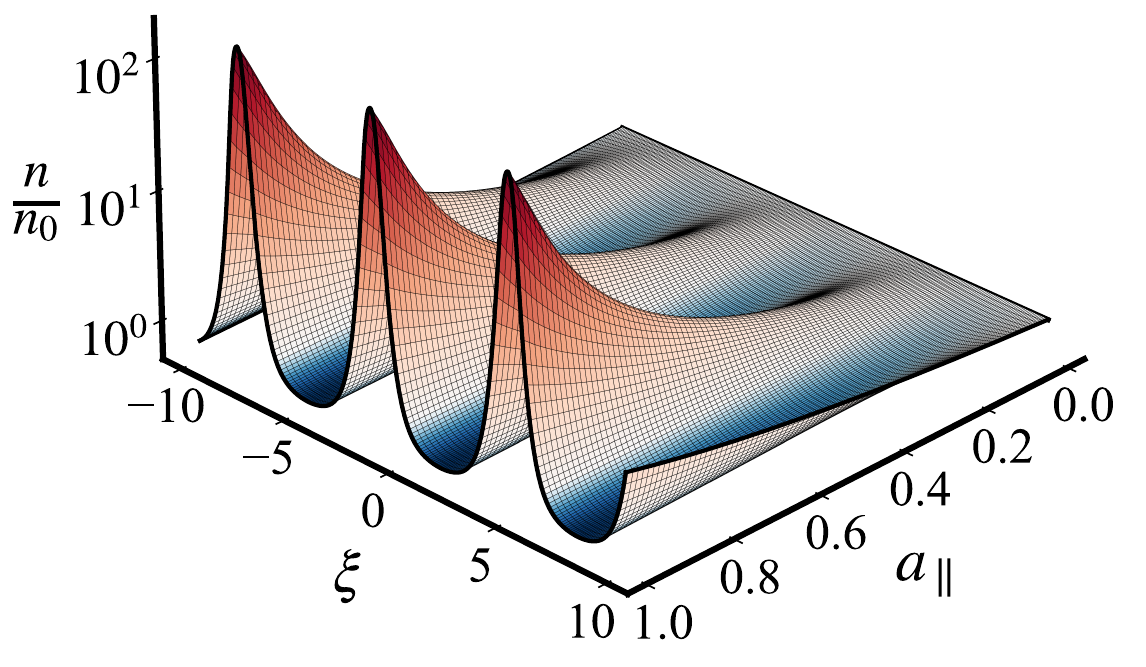}
\caption{Electron density as a function of $\alpha_\parallel$ and $\xi$, where we have used $E_\parallel(\eta) \propto \alpha_\parallel \cos(\eta)$ and $\theta \approx 0.0748$.}
\label{fig:fig2b}
\end{figure}

The continuity equation reads
\begin{equation}
\partial_t n + \partial_x (n u_e) = - k \beta c \partial_\xi n + k \partial_\xi (n u_e) = 0 ,
\label{eq:4}
\end{equation}
where $n$ is the electron density, $\partial_t n = \partial_\xi n \partial_t \xi = - k \beta c \partial_\xi n$ and $\partial_x (n u_e) = \partial_\xi (n u_e) \partial_x \xi = k \partial_\xi (n u_e) $.
The solution of equation \ref{eq:4} is
\begin{equation}
n(\xi) = \frac{\beta c n_0}{\beta c - u_e(\xi)} ,
\label{eq:6}
\end{equation}
where $n_0$ is the initial electron density. Expression \ref{eq:6} describes a modulated electron number density, seen graphically in Fig. \ref{fig:fig2b}, where for simplicity we have used $E_\parallel(\eta) \propto \alpha_\parallel \cos(\eta)$ and $\theta \approx 0.0748$.

During the boosting stage, the laser accelerates the emitted electron bunches, as seen illustratively in Fig. \ref{fig:fig1}(b) and quantitatively in Fig. \ref{fig:fig2}(d-f) at $144 \kern0.2em \mathrm{fs}$ simulation time. At that time, the electron bunches contain $\sim{2} \kern0.2em \%$ of the RP laser energy. The dashed line in Fig. \ref{fig:fig2} indicates the transition from the boosting to the collision stage. The RP laser is approximately at the focal spot, thus having flat wavefronts. The electron acceleration is quantified in Fig. \ref{fig:fig2}(e), where the electron spectrum peaks at $\sim 0.2 \pm 0.1 \kern0.2em \mathrm{GeV}$. The rest of the spectrum is approximately flat, with a cut-off energy at $\sim 2.2 \kern0.2em \mathrm{GeV}$. The longitudinal momentum component of the bunches is two orders of magnitude higher than the transverse component, a necessary condition for collimated \textgamma-photon emission.

The collision stage, illustrated in Fig. \ref{fig:fig1}(c), introduces a LP $\sim 25 \kern0.2em \mathrm{PW}$ $25 \kern0.2em \mathrm{fs}$ laser pulse, counter-propagating with respect to the electron bunches. The $0.8 \kern0.2em \mathrm{\upmu m}$ wavelength LP laser is focused in spots of $0.8-20 \kern0.2em \mathrm{\upmu m}$ at FWHM. The laser peak intensity, $I$, range is $\sim 7 \times 10^{21} - 4.4 \times 10^{24} \kern0.2em \mathrm{W cm^{-2}}$, corresponding to a dimensionless amplitude of $a_0 = e E / (m_e c \omega_l) \approx 57-1433$, where $E=\sqrt{2 I / (\varepsilon_0 c)}$ is the peak electric field, $\omega_l$ is the central laser frequency.

The laser-electron collision results in a bright \textgamma-ray flash and/or a large number of $e^{-}e^{+}$ pairs, depending on the laser focusing conditions. The \textgamma-photon emission and $e^{-}e^{+}$ pair generation ceases $\sim 50 \kern0.2em \mathrm{fs}$ after the LP laser is introduced, with an emission time of $\sim 13 \kern0.2em \mathrm{fs}$ at FWHM.

The field evolution at the end of the interaction, at $170 \kern0.2em \mathrm{fs}$ simulation time, is shown in Fig. \ref{fig:fig2}(g). The field that corresponds to the RP laser keeps altering the momentum of the electron bunches even during the interaction with the LP laser, thus enhancing the overall \textgamma-photon yield. This becomes evident by artificially removing the RP laser at the collision stage of the interaction, resulting in lower \textgamma-photon yield.

One would naturally expect the generated $e^{-}e^{+}$ pairs to be emitted along the \textgamma-photon emission direction. However, $e^{-}e^{+}$ pairs move opposite to the \textgamma-photons, as they experience acceleration by the LP laser. The generated positrons have a Maxwell-Juttner distribution peaking at $\sim 0.1 \kern0.2em \mathrm{GeV}$ and extending to $\sim 2 \kern0.2em \mathrm{GeV}$ as shown in Fig. \ref{fig:fig2}(h). The $e^{-}e^{+}$ pairs are located at symmetric sides within the laser field (Fig. \ref{fig:fig2}(i)). Their density gradually decreases due to defocusing of the laser field, but maintaining a density above the density at FWHM (for each simulation case) for several tens of femtoseconds.

%%%%%%%%%%%%%%%%%%%%%%%%%%%%%%%%%%%%%%%%%%%%%%%%%%%%%%%%%%%%%%%%%%%%%%%%%%%%
%%%%%%%%%%%%%%%%%%%%%%%%%%%%%%%%%%%%%%%%%%%%%%%%%%%%%%%%%%%%%%%%%%%%%%%%%%%%

\section*{Attosecond Gamma-Flash and Dense $e^{-}e^{+}$ Pairs}

\begin{figure*}[ht]
\centering
\includegraphics[width=1.0\linewidth]{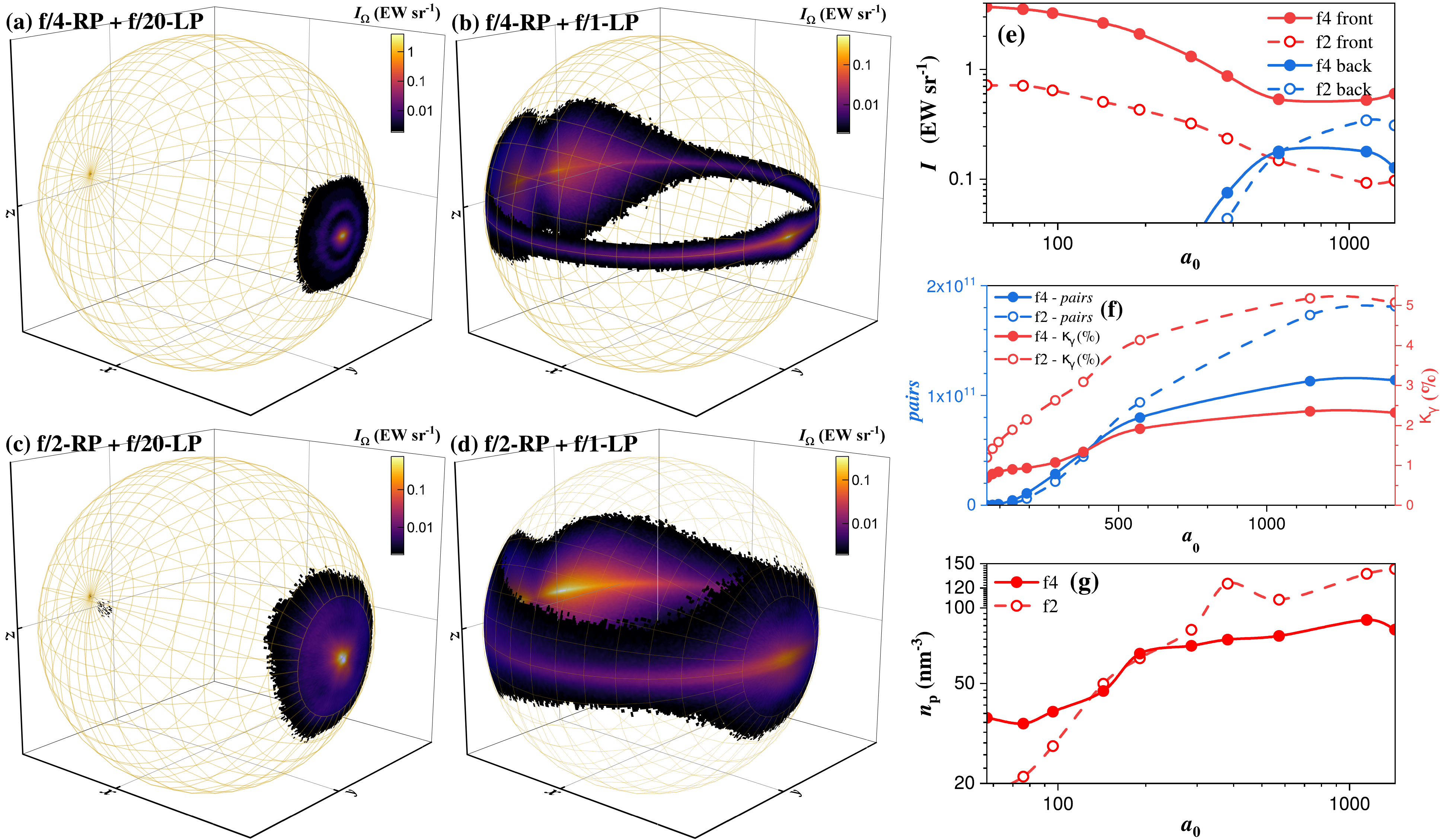}
\caption{Radiant intensity corresponding to laser combination interaction scheme of {\bf{(a)}} $\mathrm{f/4}$-RP + $\mathrm{f/20}$-LP {\bf{(b)}} $\mathrm{f/4}$-RP + $\mathrm{f/1}$-LP {\bf{(c)}} $\mathrm{f/2}$-RP + $\mathrm{f/20}$-LP {\bf{(d)}} $\mathrm{f/2}$-RP + $\mathrm{f/1}$-LP lasers. Note the one order of magnitude color-bar in Fig. \ref{fig:fig3}(a) compared to \ref{fig:fig3}(b-d). In sub-figures \ref{fig:fig3}(e-g) solid and dashed lines correspond to $\mathrm{f/4}$-RP and $\mathrm{f/2}$-RP lasers respectively. {\bf{(e)}} The peak radiant intensity versus $a_0$ for the collimated \textgamma-photon distribution (red) and for the double-lobe distribution (blue). {\bf{(f)}} The left axis (blue) shows the number of $e^{-}e^{+}$ pairs produced and the right axis (red) shows $\kappa_\gamma$ versus $a_0$. {\bf{(g)}} The maximum electron density recorded in each simulation versus $a_0$. }
\label{fig:fig3}
\end{figure*}

\begin{figure*}[ht]
\centering
\includegraphics[width=0.8\linewidth]{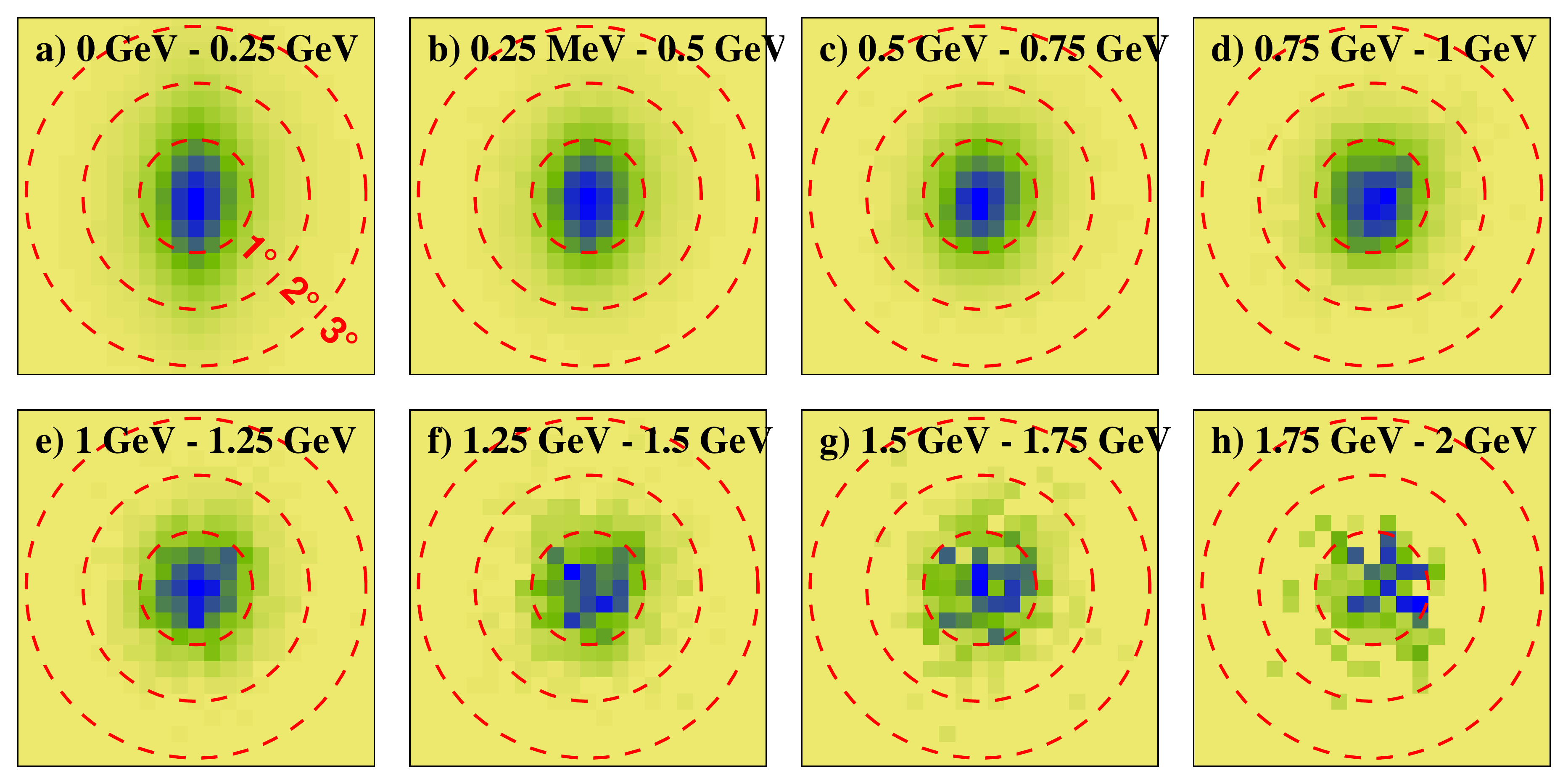}
\caption{ Normalised planar projection (on a plane $1 \kern0.2em \mathrm{m}$ far from the emitting source, the sub-figures has a width of $0.11 \kern0.2em \mathrm{m}$) of the \textgamma-photon distribution for the $\mathrm{f/20}$-LP + $\mathrm{f/4}$-RP laser case, in energy increments of $0.25 \kern0.2em \mathrm{GeV}$. The figure shows the \textgamma-beam profile dependent on the energy level. }
\label{fig:fig4}
\end{figure*}

During typical interactions of ultraintense lasers with matter, the emitter \textgamma-photon spatial distribution has a double-lobe form with a divergence of $\sim 10 \degree$ \cite{2012_NakamuraT, 2020_VyskocilJ, 2014_JiLL, 2016_StarkDJ, 2022_HadjisolomouP_b}. Although high laser to \textgamma-photon energy conversion efficiency, $\kappa_\gamma$, has been predicted by employing multi-petawatt class lasers \cite{2012_NakamuraT, 2018_LezhninKV, 2021_HadjisolomouP, 2022_HadjisolomouP, 2022_HadjisolomouP_b, 2023_HadjisolomouP_b, 2023_ShouY}, the radiant intensity, $I_\Omega$ (giving a measure of the emitted \textgamma-photon energy per unit time per unit solid angle), remains relatively low. One can increase $I_\Omega$ by reducing either the \textgamma-photon emission time or the \textgamma-photon divergence. Our proposed \textgamma-ray flash scheme treats both aforementioned aspects simultaneously.

The radiant intensity of \textgamma-photons shown in Fig. \ref{fig:fig3}(a-d). Figs. \ref{fig:fig3}(a,b) and Figs. \ref{fig:fig3}(c,d) corresponds to the use of $\mathrm{f/4}$-RP and $\mathrm{f/2}$-RP lasers as an electron driver, respectively. Figs. \ref{fig:fig3}(a,c) and Figs. \ref{fig:fig3}(b,d) correspond to the use of $\mathrm{f/20}$-LP and $\mathrm{f/1}$-LP lasers as a counter-propagating laser, respectively. For relatively low $a_0$ values of the LP laser (Figs. \ref{fig:fig3}(a,c)), a collimated \textgamma-ray flash is obtained. The \textgamma-photon beamlet is more collimated for the $\mathrm{f/4}$-RP laser case, since the ratio of the transverse to the longitudinal momentum is lower compared to the $\mathrm{f/2}$-RP laser case, although the latter gives higher electron energies. In the calculation of $I_\Omega$ we use a time interval of $1 \kern0.2em \mathrm{fs}$ (as explained later in the text). Thus, $I_\Omega$ exceeds $3.7 \kern0.2em \mathrm{EW sr^{-1}}$ in Fig. \ref{fig:fig3}(a), while it drops by a factor of $\sim 5$ in Fig. \ref{fig:fig3}(c). In this work we set the micro-wire in the centre of the RP laser, which due to laser beam pointing stability is not necessarily the case. To address this issue we performed additional simulations where the micro-wire position is shifted by $0.5 \lambda$ and $\lambda$, resulting in decrease of $I_\Omega$ by $\sim 17 \kern0.2em \%$ and $\sim 40 \kern0.2em \%$ respectively. The $I_\Omega$ versus $a_0$ is shown in Fig.  \ref{fig:fig3}(e), where the red colour represents the forward emitted \textgamma-ray flash and the blue colour represents the backward emitted double-lobe \textgamma-photon distribution. Another quantity describing the \textgamma-ray source is the radiance, $L_\Omega$, which is defined as $I_\Omega$ divided by the source size. In our case, \textgamma-photons are emitted by the localised electron bunches, which can be approximated by a circle of radius of $473 \kern0.2em \mathrm{nm}$. Thus, the radiance corresponding to Fig. \ref{fig:fig3}(a) is $\sim 5.27 \times 10^{15} \kern0.2em \mathrm{PW sr^{-1} m^{-2}}$. The highest known luminosity objects are the astrophysical \textgamma-bursts. If we assume that a \textgamma-ray burst luminosity is $10^{44} \kern0.2em \mathrm{J s^{-1}}$ \cite{2005_PiranT} and it has a photoshpere diameter diameter of $\sim 1000 \kern0.2em \mathrm{km}$ \cite{ 2021_WangXI}, then the radiance of our \textgamma-source is approaching that of \textgamma-ray bursts \cite{2010_RuffiniR, 2015_KumarP}.

By decreasing the focal spot of the LP laser from $20 \kern0.2em \mathrm{\upmu m}$ to $1 \kern0.2em \mathrm{\upmu m}$, then an $a_0 = 1146$ is reached, allowing prolific generation of $e^{-}e^{+}$ pairs. If one is interested predominantly on $e^{-}e^{+}$ pair generation, then the $\mathrm{f/2}$-RP laser case (injection and boosting stage) has a higher yield, since electrons reach higher energies. Approximately $1.8 \times 10^{11}$ $e^{-}e^{+}$ pairs are generated by the LP laser with the electron bunches. The number of generated $e^{-}e^{+}$ pairs versus $a_0$ is shown in the left axis of Fig. \ref{fig:fig3}(f), while the right axis shows $\kappa_\gamma$. The figure indicates an otherwise obvious conclusion, that the more \textgamma-photons are emitted, the more $e^{-}e^{+}$ pairs are generated. Since those $e^{-}e^{+}$ pairs are generated in the small volume where the electron bunches are confined, their density value can exceed that of lithium, as seen in Fig. \ref{fig:fig3}(g). The $e^{-}e^{+}$ pairs are driven by the LP laser in its propagation direction. Those backward moving fast electrons/positrons move in a strong laser field and emit secondary \textgamma-photons with a double-lobe pattern. As seen in Fig. \ref{fig:fig3}(e), $I_\Omega$ of the secondary emitted \textgamma-photons can exceed that of the forward \textgamma-photons. The forward \textgamma-photon distribution decreases due to the $e^{-}e^{+}$ pair generation. We have proven that by reducing the focal spot of the LP laser from $20 \kern0.2em \mathrm{\upmu m}$ to $4 \kern0.2em \mathrm{\upmu m}$ and dropping the power of the RP laser from $25 \kern0.2em \mathrm{PW}$ to $1 \kern0.2em \mathrm{PW}$, then the resulting \textgamma-photon spectrum does not change significantly.

\begin{figure*}[ht]
\begin{subfigure}{0.33\textwidth}
  \centering
  \includegraphics[width=1.0\linewidth]{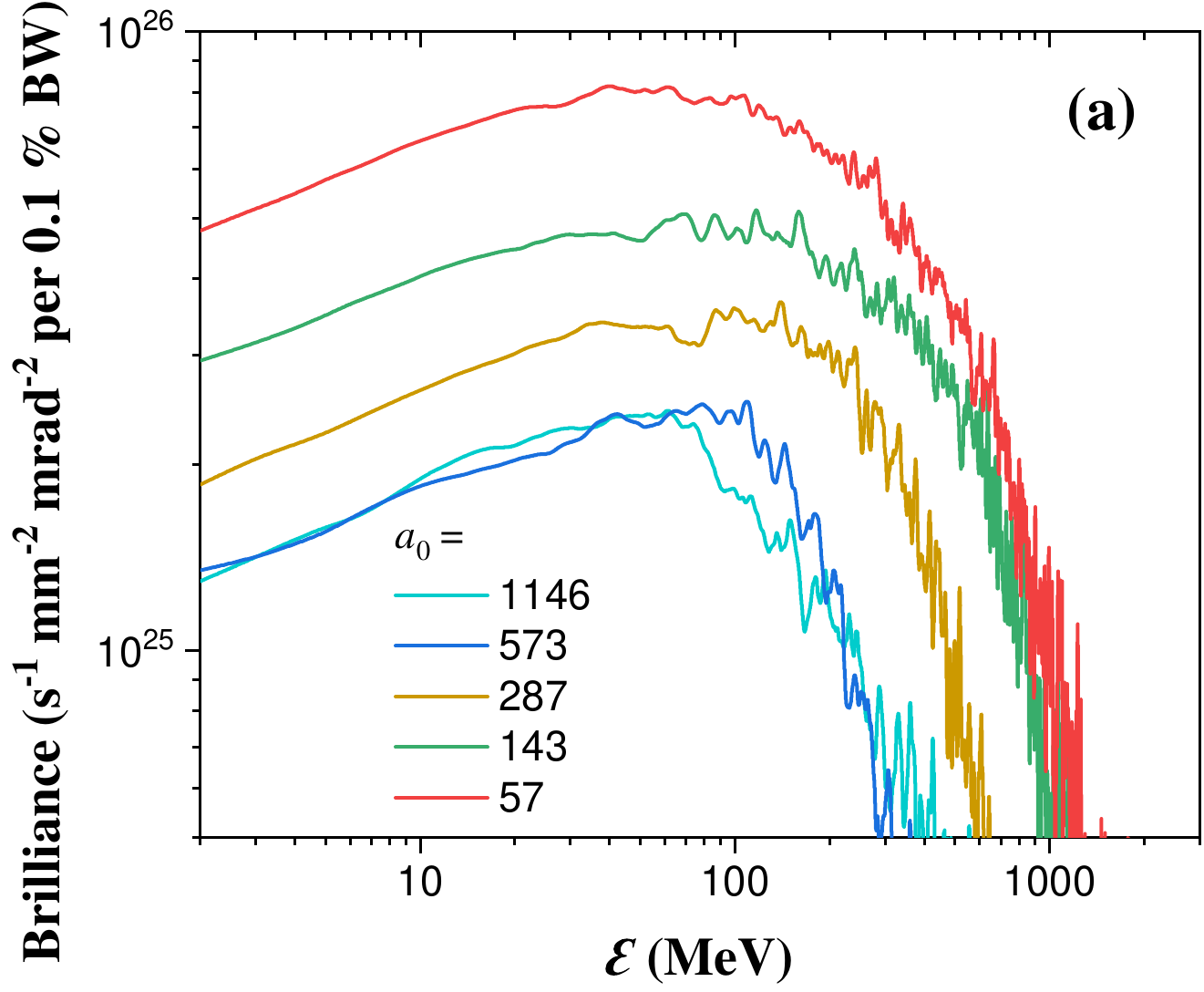}
\end{subfigure}
\begin{subfigure}{0.33\textwidth}
  \centering
  \includegraphics[width=1.0\linewidth]{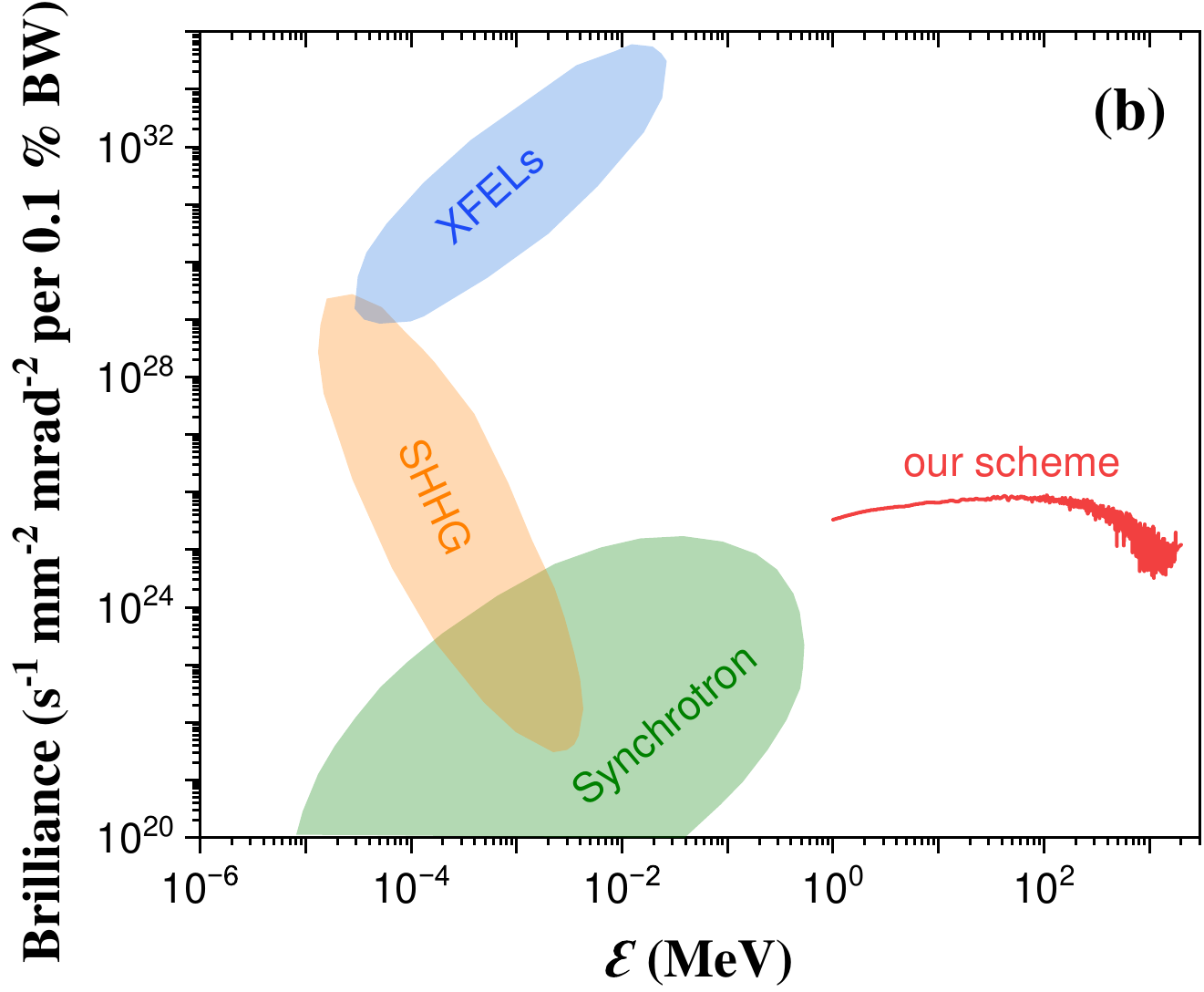}
\end{subfigure}
\begin{subfigure}{0.315\textwidth}
  \centering
  \includegraphics[width=1.0\linewidth]{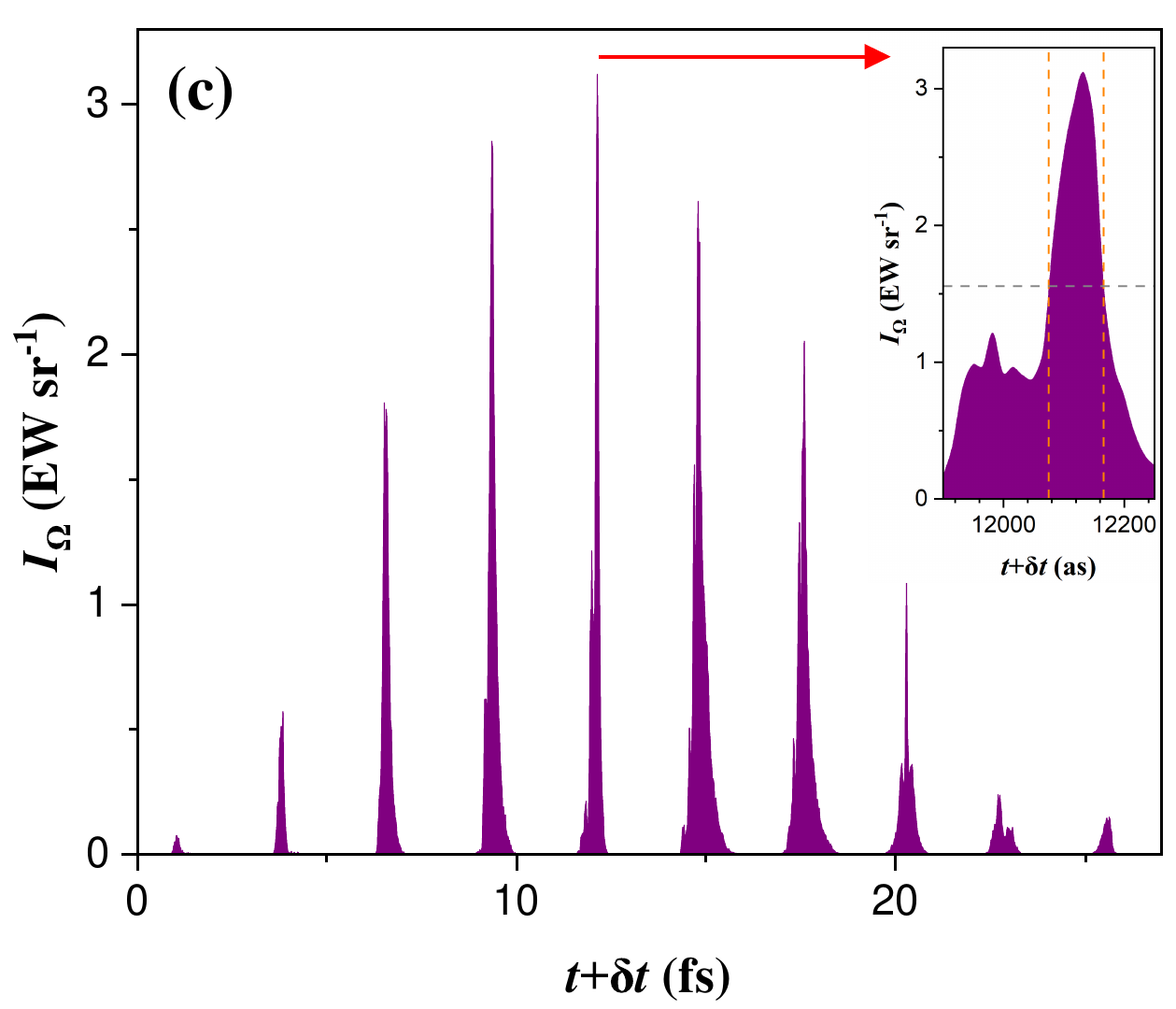}
\end{subfigure}
\caption{ {\bf{(a)}} Brilliance as a function of $a_0$, corresponding the $\mathrm{f/20}$-LP + $\mathrm{f/4}$-RP laser case. {\bf{(b)}} Brilliance of x-ray sources from synchrotrons (green), SHHG (orange) and XFELs (blue) (see references \cite{2019_JahnO, 2023_SchwartzCP}) compared to our scheme (red). {\bf{(c)}} Temporal profile of the \textgamma-ray flash, revealing a series of attosecond \textgamma-photon pulses. The temporal axis employs an offset to shift the distribution at axis origin. The inset graph zooms at the \textgamma-photon pulse at $\sim 10 \kern0.2em \mathrm{fs}$, indicated by the red arrow.}
\label{fig:fig5}
\end{figure*}

The small divergence on the \textgamma-ray flash shown in Figs. \ref{fig:fig3}(a,c) is better realised in Fig. \ref{fig:fig4}. The sub-figures show projection of \textgamma-photons on a plane orthogonal to the laser propagation axis, for energy intervals increasing by $0.25 \kern0.2em \mathrm{GeV}$. The red circles correspond to divergence increments of $1 \degree$. The \textgamma-photon counts are normalised to the peak value in each sub-figure.

The figure demonstrates a collimated \textgamma-ray flash with a slightly elliptical profile. For notation, we assume an ellipse of major and minor axis corresponding to full-angle divergence defined as $(a,b)$; the major axis coincides with the LP laser electric field oscillation direction. Here, $(a,b) \approx (1.95 \degree, 1.45 \degree)$ for the cumulative \textgamma-photon signal, dominated by the relatively low ($< 0.25 \kern0.2em \mathrm{GeV}$) energy \textgamma-photons. The \textgamma-photon spatial distribution becomes more narrow for $0.5-0.75 \kern0.2em \mathrm{GeV}$, obtaining $(a,b) \approx (1.25 \degree, 1 \degree)$. For the $\mathrm{f/20}$-LP + $\mathrm{f/4}$-RP laser case, the \textgamma-photon divergence increases to $\sim 4 \degree$, with an asymmetric distribution.

For applications, one needs to know not only how intense the \textgamma-ray flash is in a specific solid angle, but also information for the \textgamma-photon energy. The aforementioned information are obtained through brilliance, as shown in Fig. \ref{fig:fig5}(a) for the $\mathrm{f/4}$-RP laser case and for various $a_0$ cases of the LP laser. The highest brilliance is achieved by a relatively weakly focused laser to a $20 \kern0.2em \mathrm{\upmu m}$ diameter spot. The brilliance does not follow the typical exponentially decaying pattern which peaks for lower \textgamma-photon energies \cite{2016_ZhuXL, 2018_ZhuXL, 2018_WangWM, 2020_ZhuXL, 2022_HadjisolomouP_b}, rather the peak distribution is at $\sim 50 \kern0.2em \mathrm{MeV}$. This peak value conveniently matches photonuclear reactions \cite{2022_SunXY, 2022_DeievOS}, which require tens of MeV energies. Our \textgamma-photon source provides an extremely bright source in the multi-MeV range, reaching $\sim 9 \times 10^{25} \kern0.2em \mathrm{ s^{-1} mm^{-2} \kern0.2em per \kern0.2em 0.1\% \kern0.2em BW }$, corresponding to a total number of \textgamma-photons (with energy above $1 \kern0.2em \mathrm{MeV}$) of $\sim 9.6 \times 10^{11}$. In Fig. \ref{fig:fig5}(b) we compare our \textgamma-photon source with other schemes (see \cite{2019_JahnO} and references therein), namely synchrotron, surface high-order harmonic generation (SHHG) and x-ray free electron lasers (XFELs). The proposed scheme generates high brilliance \textgamma-photon pulses at energies which cannot be achieved by the aforementioned schemes.

Temporal analysis of the \textgamma-ray flash for the peak brilliance case is shown in Fig. \ref{fig:fig5}(c). The \textgamma-photons are emitted in a series of attosecond pulses, with a period matching the laser frequency. The inlet figure zooms in the more intense of those pulses, having a duration of $\sim 90 \kern0.2em \mathrm{as}$ at FWHM. Since the envelope of the \textgamma-photon emission profile is $\sim 13 \kern0.2em \mathrm{fs}$ at FWHM, the cumulative \textgamma-photon emission time is approximately $\sim 1 \kern0.2em \mathrm{fs}$, as used above in $I_\Omega$ calculation. This is verified by Fig. \ref{fig:fig5}(c) where the interval time is used, giving a similar value for the peak $I_\Omega$ compared to Fig. \ref{fig:fig3}(a).

%%%%%%%%%%%%%%%%%%%%%%%%%%%%%%%%%%%%%%%%%%%%%%%%%%%%%%%%%%%%%%%%%%%%%%%%%%%%
%%%%%%%%%%%%%%%%%%%%%%%%%%%%%%%%%%%%%%%%%%%%%%%%%%%%%%%%%%%%%%%%%%%%%%%%%%%%

\section*{Conclusions}

In this paper we introduce a novel dyadic laser-matter interaction scheme for emission of highly collimated, ultrabright, attosecond \textgamma-ray flashes and generation of dense $e^{-}e^{+}$ pairs. The parameters of our work correspond to the dual-beam $25 \kern0.2em \mathrm{PW}$ EP-OPAL laser currently in preparatory phase in the University of Rochester, USA. Our interaction scheme employs a RP laser interacting with a micro-wire target, emitting localised electron bunches with longitudinal momentum component significantly higher than the transverse one. A counter-propagating LP laser interacts with the electron bunches, emitting a collimated \textgamma-photon beamlet. The divergence value is a function of the emitted \textgamma-photon energy interval and it reaches values as low as $\sim 1 \degree$. Our scheme offers an ultrabright \textgamma-photon source of $\sim 9 \times 10^{25} \kern0.2em \mathrm{ s^{-1} mm^{-2} \kern0.2em per \kern0.2em 0.1\% \kern0.2em BW }$ at $\sim 50 \kern0.2em \mathrm{MeV}$. By controlling the LP laser focusing conditions we achieve prolific $e^{-}e^{+}$ pair generation, while at times of $e^{-}e^{+}$ pair peak generation rate, the positron density reaches a value that exceeds the solid density level.

The interaction of matter with electromagnetic field occurs in astrophysical scales, where \textgamma-photon emission has been observed in the form of \textgamma-ray bursts \cite{2001_RuffiniR, 2001_RuffiniRb, 2019_AmiriM, 2020_MarcoteB}. Moreover, the unique plasma features achieved by our setup in comparison to other schemes falls into the strong-field quantum electrodynamics environment where $e^{-}e^{+}$ pairs exhibit collective effects. Quantum electrodynamics plasmas lay on a strong theoretical framework, of which the experimental verification faces challenges on how to achieve such a dense $e^{-}e^{+}$ pair plasma \cite{2020_ZhangP, 2023_ChenH, 2024_ArrowsmithCD}. Such dense plasmas can occur in the viscinity of neutron stars \cite{2022_PhilippovA} and/or black holes \cite{2016_BambiC} and have recently been achieved in the laboratory with an $e^{-}e^{+}$ pair density of $\sim 10^{18} \kern0.2em \mathrm{m^{-3}}$ with the aim of testing the microphysics of astrophysical observations \cite{2024_ArrowsmithCD}. We hope our proposed scheme, predicting an $e^{-}e^{+}$ pair density more than ten orders of magnitude higher, to pave the way on generation of such plasma environments and illuminate novel insights into fundamental physics.

Temporal analysis of the emitted \textgamma-photon distribution reveals a series of $\sim 100 \kern0.2em \mathrm{as}$ pulses, with a high radiant intensity of $\sim 3 \kern0.2em \mathrm{EW sr^{-1}}$. Compared with the pioneering work of Zewail, Mourou, L'Huillier, Agostini, and Krausz that has transformed chemistry using femtosecond and attosecond optical pulses (see \cite{2006_MourouGA, 2009_KrauszF} and references therein) and Hajdu and Chapman’s invention using X-ray pulses, which brought the atomic view of materials \cite{2000_NeutzeR, 2002_HajduJ, 2006_ChapmanHN, 2007_ChapmanHN}, our invention of attosecond gamma-ray pulses can revolutionize photo-nuclear physics creating new era of science and technology.

%%%%%%%%%%%%%%%%%%%%%%%%%%%%%%%%%%%%%%%%%%%%%%%%%%%%%%%%%%%%%%%%%%%%%%%%%%%%

\section*{Methods}

The results of this work were obtained through use of the quantum electrodynamics \cite{2012_RidgersCP, 2013_RidgersCP, 2014_RidgersCP} 3D EPOCH \cite{2015_ArberTD} PIC code. The Higuera-Cary \cite{2017_HigueraAV} particle pusher is employed instead of the default Boris option, to obtain a more accurate trajectory of the relativistic electrons present in the simulation. The overall simulation results from two dependent simulations. In the first simulation we calculate the RP fields externally, and then import them into EPOCH, covering the left half of the simulation box. The laser pulse has a $0.8 \kern0.2em \mathrm{\upmu m}$ wavelength, $25 \kern0.2em \mathrm{fs}$ pulse duration and it corresponds to a $25 \kern0.2em \mathrm{PW}$ laser. The laser is assumed to be focused by either $\mathrm{f/2}$ or $\mathrm{f/4}$ parabolas, resulting in an intensity profile of $I=I_0 r^2 \exp(-r^2/r_0^2)$, where $I_0 \approx 4.7 \times 10^{23} \kern0.2em \mathrm{W cm^{-2}}$ and $r_0 \approx 1.2 \kern0.2em \mathrm{\upmu m}$ for the $\mathrm{f/2}$ case and $I_0 \approx 1.2 \times 10^{23} \kern0.2em \mathrm{W cm^{-2}}$ and $r_0 \approx 2.4 \kern0.2em \mathrm{\upmu m}$ for the $\mathrm{f/4}$ case.. The right half of the box contains no fields; there, a lithium cylinder is placed along the laser propagation axis, with radius of $\lambda$ and length of $4 \lambda$. A moving window moving with the speed of light is used in the simulations, starting at $14.97 \kern0.2em \mathrm{fs}$ and stops at $27.35 \kern0.2em \mathrm{fs}$ for the $\mathrm{f/2}$ case, while for the $\mathrm{f/4}$ case the window starts at $4.23 \kern0.2em \mathrm{fs}$ and stops at $117.53 \kern0.2em \mathrm{fs}$.

The initial and final time of the moving window are chosen in such a way that the simulation stops at a time of $2 \sigma$ (where $\sigma$ is the standard deviation of the pulse temporal profile) prior the electron bunches reach their highest energy. The field and particle data from the first simulation are imported into a second simulation, where a LP laser (same wavelength, temporal profile and power as the first laser) is launched from the right boundary of the simulation box, with a temporal offset of $2 \sigma$, focusing at a distance of $2 \sigma c$ from the boundary. As a result, the peak intensity of the linearly polarised laser meets the electrons (the centre of their distribution) when they reach their peak energy value. The LP laser focal spot ranges from $0.8 \kern0.2em \mathrm{\upmu m}$ to $20 \kern0.2em \mathrm{\upmu m}$. In all stages the simulation box has dimensions of $15.36 \kern0.2em \mathrm{\upmu m} \times 20.48 \kern0.2em \mathrm{\upmu m} 20.48 \kern0.2em \mathrm{\upmu m}$ with a cell resolution of $5 \kern0.2em \mathrm{nm} \times 40 \kern0.2em \mathrm{nm} 40 \kern0.2em \mathrm{nm}$.

%%%%%%%%%%%%%%%%%%%%%%%%%%%%%%%%%%%%%%%%%%%%%%%%%%%%%%%%%%%%%%%%%%%%%%%%%%%%

\section*{Data availability}
The data that support the findings of this study are available from the corresponding author, PH, upon reasonable request.

\bibliography{biblio_2024CP}

\section*{Acknowledgements}
\par The authors thank Prof. Hajdu for useful discussion. The EPOCH code is in part funded by the UK EPSRC grants EP/G054950/1, EP/G056803/1, EP/G055165/1 and EP/M022463/1.

\section*{Author contributions}
Author contributions: P.H., T.M.J. and S.V.B. conceived the work and participated in the interpretation of results; P.H. performed the PIC simulations and analysed the data;; P.H. and P.V. prepared the figures; R.S., A.J.M. and C.P.R. provided theoretical support for the results. The manuscript was written by all authors.

\section*{Competing interests}
The authors declare no competing interests.

\end{document}